\documentclass[12pt]{article}

\usepackage{graphicx}
\usepackage{float}
\usepackage{mathtools}
\usepackage{amssymb}
\usepackage{slashed}
\usepackage{subcaption}
\usepackage{placeins}
\usepackage[bottom]{footmisc}

\usepackage{listings}
\usepackage{lineno}

\usepackage{feynmp}
\usepackage{hyperref}

\usepackage[amssymb]{SIunits}

\newcommand\pubnumber{DPF2015-359}
\newcommand\pubdate{November 1, 2015}

\def\napoli{$^1$Department of Physics, Royal Holloway, University of London, Egham Hill, Egham, Surrey TW20 0EX, UK}

\def\Title#1{\begin{center} {\Large #1 } \end{center}}
\def\Author#1{\begin{center}{ \sc #1} \end{center}}
\def\Address#1{\begin{center}{ \it #1} \end{center}}

\newcommand\pubblock{\rightline{\begin{tabular}{l} \pubnumber\\
         \pubdate  \end{tabular}}}
\newenvironment{Abstract}{\begin{quotation}  }{\end{quotation}}
\newenvironment{Presented}{\begin{quotation} \begin{center}
             PRESENTED AT\end{center}\bigskip
      \begin{center}\begin{large}}{\end{large}\end{center} \end{quotation}}

\def\beq{\begin{equation}}
\def\eeq#1{\label{#1}\end{equation}}
\def\eeqn{\end{equation}}

\def\beqa{\begin{eqnarray}}
\def\eeqa#1{\label{#1}\end{eqnarray}}
\def\eeqan{\end{eqnarray}}

\let\bar=\overbar

\def\Dslash{\not{\hbox{\kern-4pt $D$}}}
\def\dslash{\not{\hbox{\kern-2pt $\del$}}}

\def\msb{{\bar{\ssstyle M \kern -1pt S}}}

\begin{document}
\begin{titlepage}
\pubblock

\vfill
\Title{The Performance and Development of the Inner Detector Trigger Algorithms at ATLAS for LHC Run 2}
\vfill
\Author{Benjamin Sowden$^1$ on behalf of the ATLAS collaboration}
\Address{\napoli}
\vfill
\begin{Abstract}
The upgrade to the ATLAS trigger for LHC Run 2 is presented including a description of the design and performance of the newly reimplemented tracking algorithms. The profiling infrastructure, constructed to provide prompt feedback from the optimisation procedure is described including the methods used to monitor the relative performance improvements as the code evolves. The performance of the trigger on the first data collected in the LHC Run 2 is presented.
\end{Abstract}
\vfill
\begin{Presented}
DPF 2015\\
The Meeting of the American Physical Society\\
Division of Particles and Fields\\
Ann Arbor, Michigan, August 4--8, 2015\\
\end{Presented}
\vfill
\end{titlepage}
\def\thefootnote{\fnsymbol{footnote}}
\FloatBarrier
\section{Introduction}

The ATLAS detector \cite{Aad:2008zzm} is a large general purpose particle detector situated on the Large Hadron Collider (LHC) \cite{Evans:2008zzb} in Geneva, Switzerland.
ATLAS successfully collected $\sim\unit{25}{\reciprocal{\femto\barn}}$ of $7$ and $\unit{8}{\tera\electronvolt}$ proton-proton collision data between 2010 and 2012, this period of running is known as Run 1.
From early 2013 to early 2015 the LHC and experiments were in a long shutdown during which they underwent maintenance and upgrade work.
Run 2 started this year with the first $\unit{13}{\tera\electronvolt}$ collisions being recorded.

The data taking conditions have changed significantly from Run 1 to Run 2.
As mentioned there is a large increase in centre of mass energy from Run 1.
Furthermore peak luminosity and the maximum number of interactions per bunch crossing (pileup) will increase ($\unit{7\times10^{33}}{\rpsquare{\centi\metre}\usk\reciprocal\second}$ to $\unit{2\times10^{34}}{\rpsquare{\centi\metre}\usk\reciprocal\second}$ and $40$ to $50-55$ respectively) later in Run 2 when the LHC has ramped to peak performance.
The spacing between bunch crossings will be reduced from \unit{50}{\nano\second} to \unit{25}{\nano\second}, doubling the collision rate from \unit{20}{\mega\hertz} to \unit{40}{\mega\hertz}.

This change in conditions will cause a dramatic increase in the rate of processes which pass the trigger with some of the processes which contribute most to the rate of the first stage of the trigger seeing a factor of five increase.
To handle this increased rate while maintaining optimal selection efficiency the ATLAS detector and trigger system have been upgraded.
The major detector upgrade influencing the inner detector trigger is the addition of a fourth pixel layer closer to the beamline called the Insertable B Layer (IBL) \cite{Capeans2010}.
The IBL improves vertexing performance and impact parameter resolution, and adds robustness against missed hits and disabled modules in track identification.
The High Level Trigger (HLT) architecture has been simplified such that both the Run 1 software stages of the HLT (the Level 2 (L2) and Event Filter (EF)) have been merged into a single process, running on a single HLT computing cluster node.
Also a new hardware-based track preprocessor (FTK) \cite{Shochet2013} is planned to be added early in Run 2.
It will process events after the Level 1 (LVL1) hardware-based trigger accept in order to seed the HLT algorithms.

\FloatBarrier
\section{The Inner Detector Trigger}

\begin{figure}[htb]
  \centering
  \includegraphics[width=0.7\textwidth]{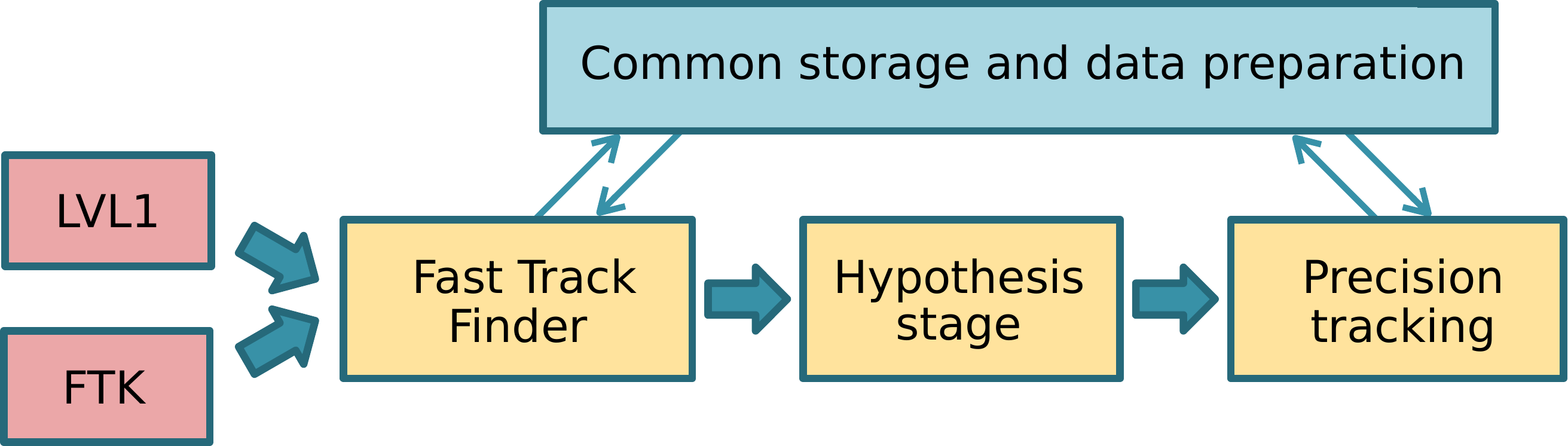}
  \caption{Layout of the inner detector tracking trigger in Run 2. Note the FTK is due to be added in 2016.}
  \label{fig:IDTrig}
\end{figure}

The inner detector trigger software has been rewritten to take advantage of the new HLT trigger framework.
The same two step structure is retained as can be seen in Fig.~\ref{fig:IDTrig}.
The first step, as with Run 1, is a custom written fast tracking algorithm which performs pattern recognition and tracking on the hit data\footnote{Hit data are silicon hits in the pixel and strip detectors, plus hits in the Transition Radiation Tracker.}.
This step has been completely re-written with the FTK considered from the start, meaning tracks provided by the FTK can be integrated seamlessly into the system once the FTK is commissioned.
This produces a hypothesis for the event which is passed to and used as a starting point for the second step.
The second step is precision tracking, which utilises an optimised subset of the tracking algorithms used offline.
This second step is slower than the first but does a more thorough job of identifying the objects constructed using the inner detector tracks (e.g. electron, muons, etc.).
The new single node running allows for the two stages of the trigger to share the data preparation so detector information only needs to be read out once.
Additionally, a single data format is used by both stages.
The seeding of the precision tracking from the fast track finder allows for the time-consuming pattern finding stage of the trigger algorithm to not be repeated.
\FloatBarrier
\section{Profiling and Optimisation}

There are two main factors that contribute to the substantially reduced run time of the HLT inner detector tracking algorithms: improvements made in the offline tracking algorithms and the restructuring of the trigger.

\begin{figure}[htb]
  \centering
  \begin{subfigure}[htb]{0.47\textwidth}
      \centering
      \includegraphics[width=\textwidth]{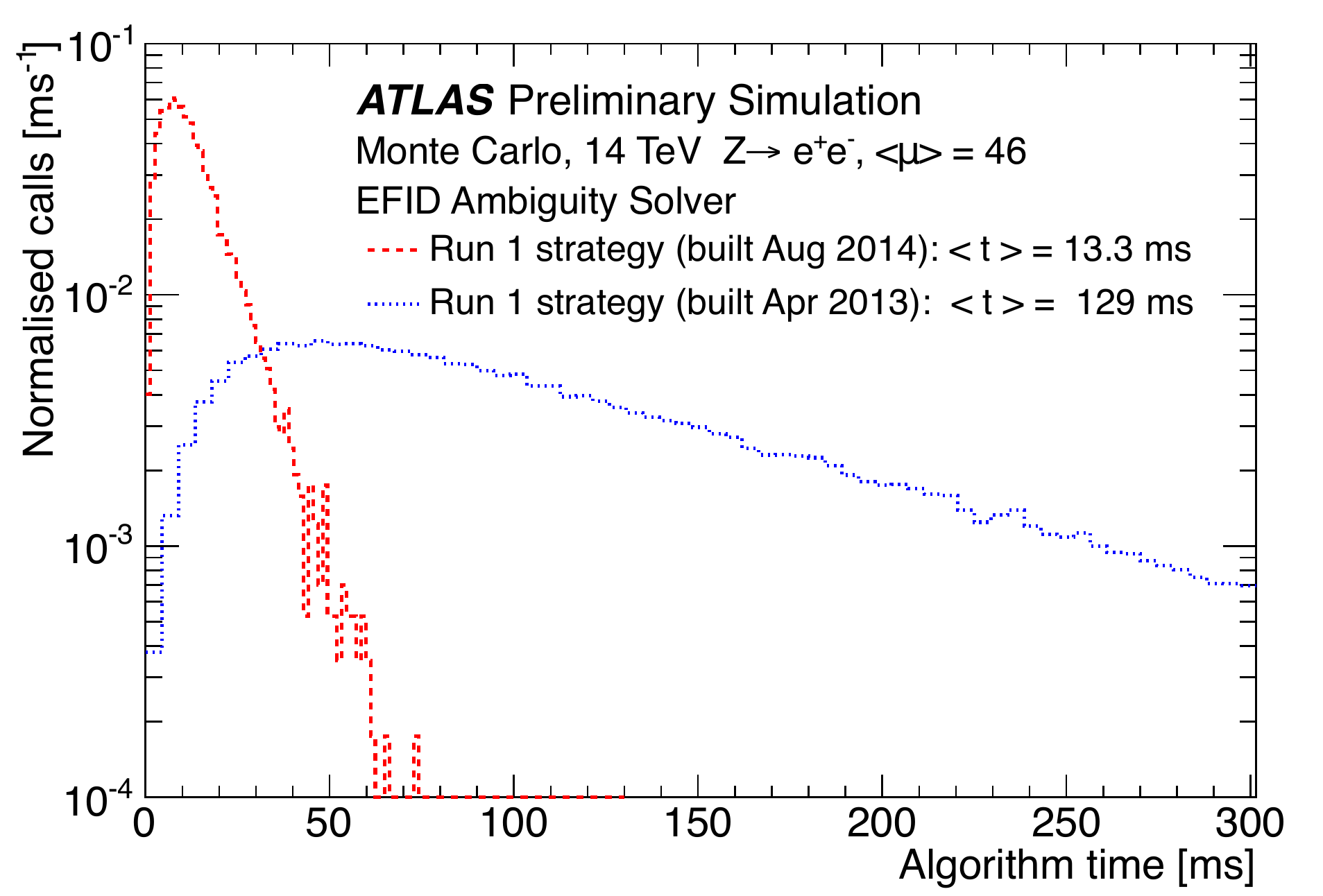}
      \caption{}
      \label{fig:Run1TriggerTimes:AS}
  \end{subfigure}
  ~
  \begin{subfigure}[htb]{0.47\textwidth}
      \centering
      \includegraphics[width=\textwidth]{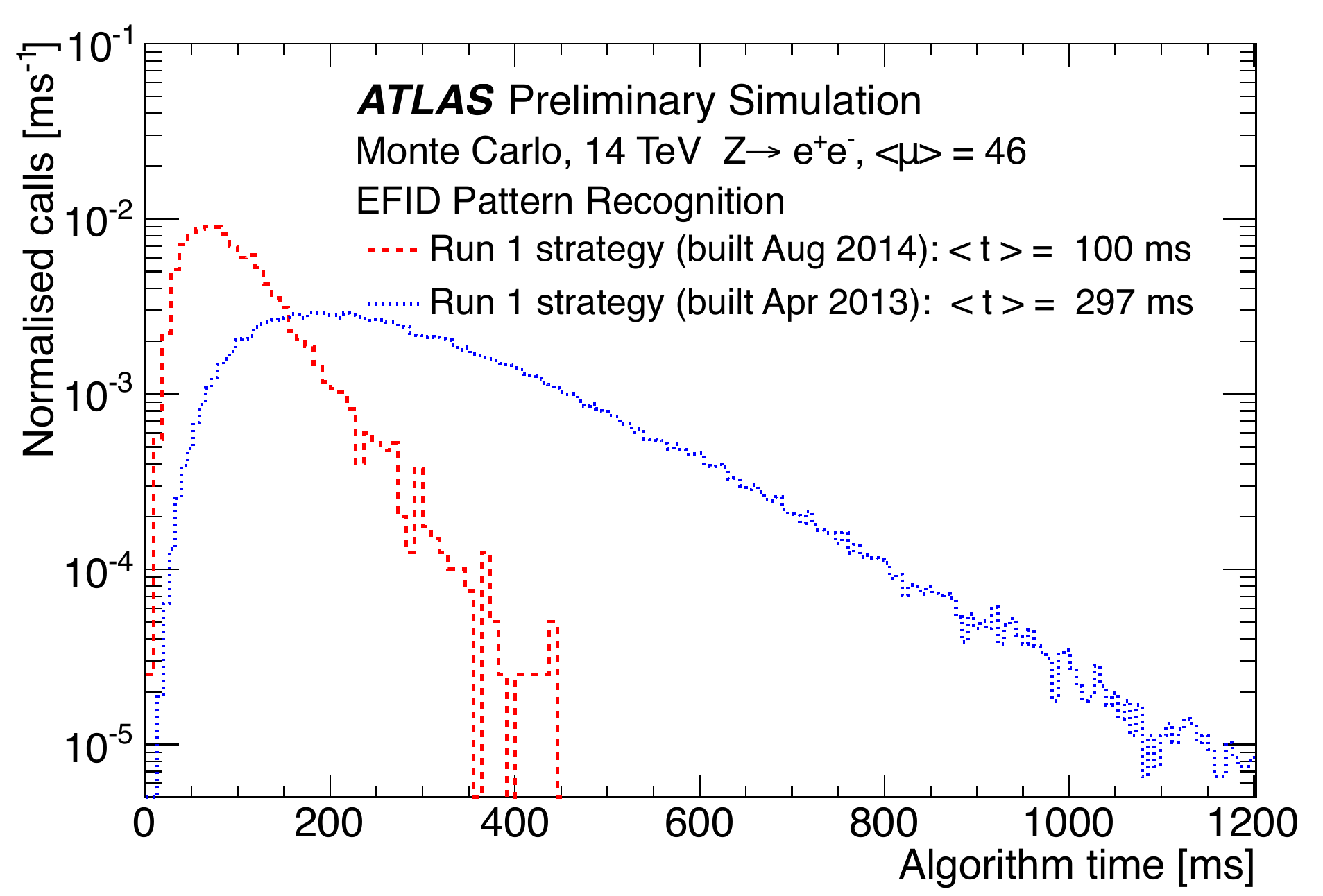}
      \caption{}
      \label{fig:Run1TriggerTimes:PR}
  \end{subfigure}
  \caption{The distribution of processing times per call for (a) the ambiguity solver and (b) the pattern recognition algorithms within the Event Filter Inner Detector (EFID) trigger tracking on simulated Z$\rightarrow$ee data with mean pileup of 46. Shown are the times for the tracking strategy used during the ATLAS Run 1 data taking in 2012 but implemented in different versions of the ATLAS code; one built in April 2013 and one more recently built in August 2014. The later is significantly faster as the offline tracking algorithms used in the Event Filter tracking underwent significant optimisation. Taken from Ref.~\cite{IDTrigTimeEff}.}
  \label{fig:Run1TriggerTimes}
\end{figure}

Firstly large gains in speed are made by incorporating improvements from the offline tracking algorithms \cite{OfflineTime} into the precision tracking section of the trigger which uses slightly modified versions of these algorithms.
The improvements lead to a factor three reduction in time the full offline reconstruction.
A large part of these gains are achieved in the two most time-consuming parts of the precision tracking algorithm\footnote{Together they constitute nearly $90\%$ of the ID tracking time at the Event Filter}, as shown in Fig.~\ref{fig:Run1TriggerTimes}.
The two trigger algorithms shown are the ambiguity solver and the pattern recognition.
The pattern recognition searches for potential tracks in hit data, after which the ambiguity solver resolves ambiguities between near-by and crossing tracks.

\begin{figure}[htb]
  \centering
  \begin{subfigure}[htb]{0.47\textwidth}
      \centering
      \includegraphics[width=\textwidth]{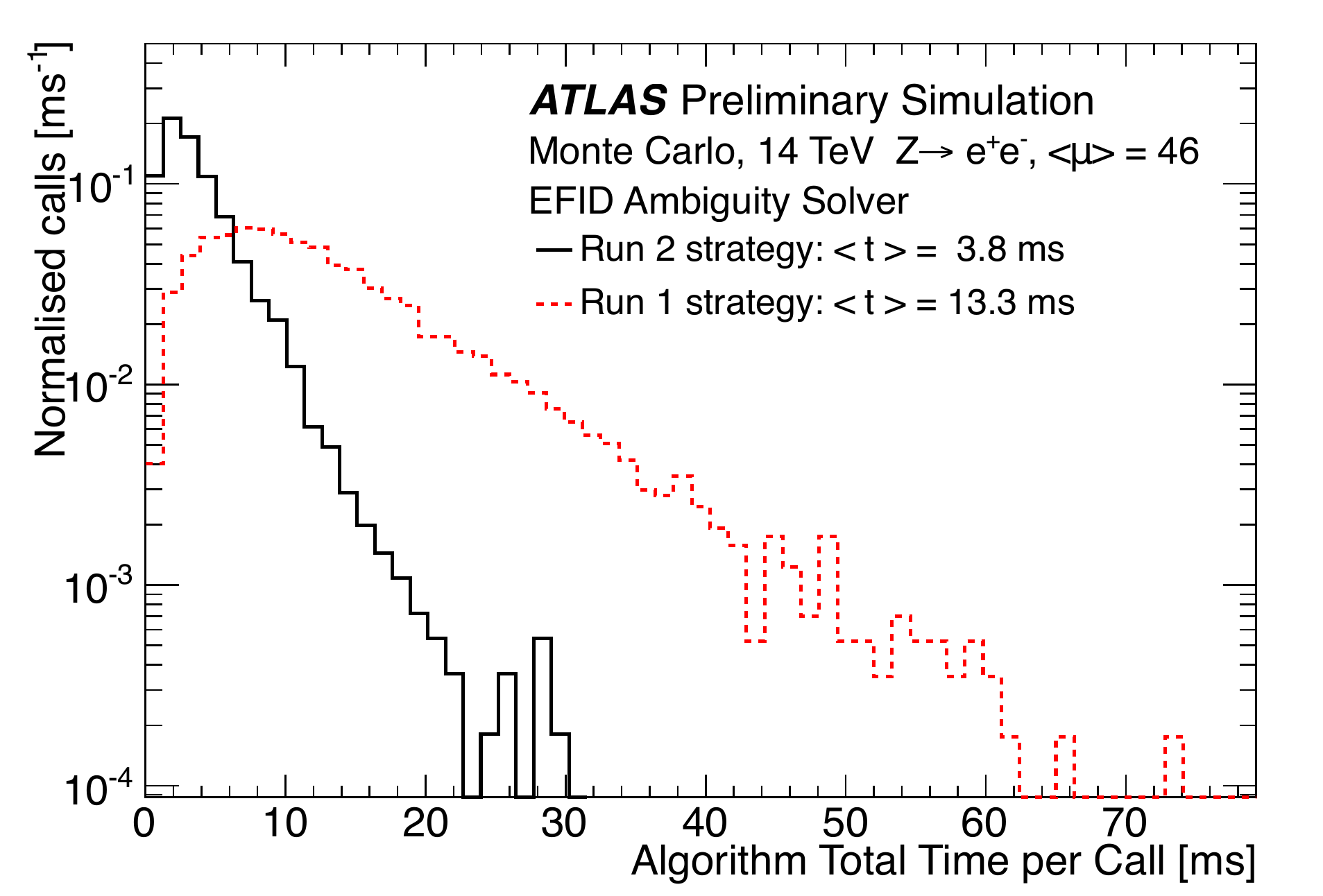}
      \caption{}
  \end{subfigure}
  ~
  \begin{subfigure}[htb]{0.47\textwidth}
      \centering
      \includegraphics[width=\textwidth]{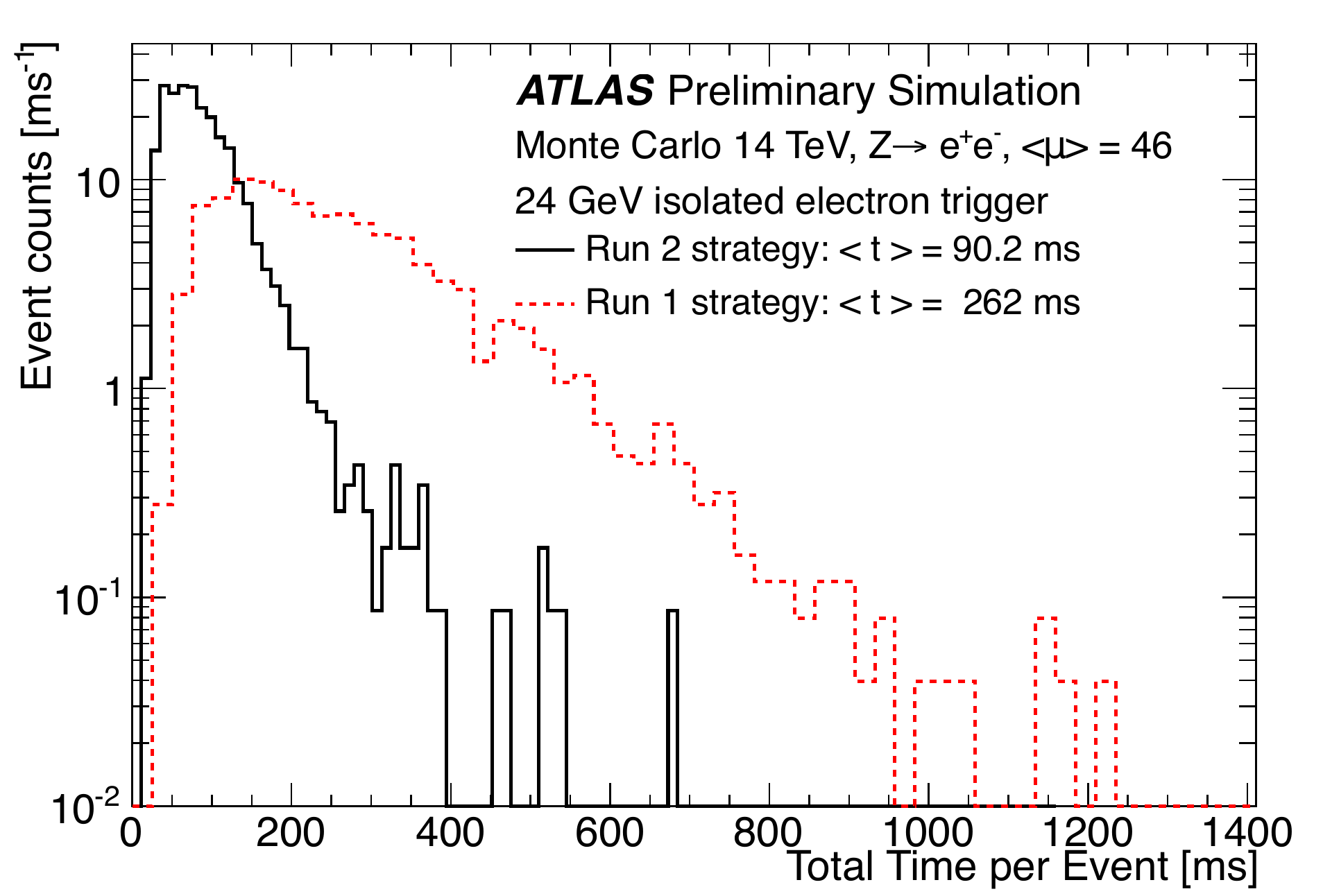}
      \caption{}
  \end{subfigure}
  \caption{The distribution of (a) the processing times per call for the ambiguity-solver algorithm and (b) the time per event for the full execution of the electron trigger on simulated Z$\rightarrow$ee data with mean pileup of 46. Shown are the comparison of timings for two alternative ``strategies'' for the ID Trigger; The ``Run 1 strategy'' runs the improved offline algorithms considered in Fig.~\ref{fig:Run1TriggerTimes}. The ``Run 2 strategy'' runs the newly structured trigger discussed abovee. Taken from Ref.~\cite{IDTrigTimeEff}.}
  \label{fig:Run2TriggerTimes}
\end{figure}

\pagebreak
The second significant speed-up is due to the restructuring of the trigger.
As the fast track finder's result is reused by the precision tracking, the Run 1 Event Filter pattern recognition stage can be skipped entirely.
Fig.~\ref{fig:Run2TriggerTimes} shows the improvements in processing time due only to this change, for both the ambiguity solver algorithm and the full chain of algorithms.
The detailed tracking can now run on a reduced number of patterns, selected in the previous stage, thus reducing its load.
The electron trigger processing time on simulated data\footnote{The full trigger processing times per event include the time spent running the chain multiple times in events with more than one electron candidate.
It should be noted that the total trigger processing includes the time for the calorimeter reconstruction and additional, non-tracking, algorithms, which are common to both strategies and contribute approximately 22 ms to the total event processing time in each case.} has seen a substantial decrease after incorporating all of the above improvements with the average run time being almost a factor of three less than the Run 1 strategy.

\FloatBarrier
\section{Performance on First Data}

The performance of the triggers was assessed with \unit{13}{\tera\electronvolt} data collected in July 2015 by the ATLAS detector.
This data was collected using dedicated performance triggers which selected events regardless of results from the inner detector trigger processing.
Efficiencies, residuals and resolutions are calculated relative to the tracks found by the offline reconstruction software.

Only good quality offline tracks with at least 2 pixel clusters and 6 silicon strip clusters are used, the tracks are also required to be in the region corresponding to the inner detector acceptance (with absolute pseudorapidity\footnote{ATLAS uses a right-handed coordinate system with its origin at the nominal interaction point (IP) in the centre of the detector and the $z$-axis along the beam pipe. The $x$-axis points from the IP to the centre of the LHC ring, and the $y$-axis points upward. Cylindrical coordinates $(r,\phi)$ are used in the transverse plane, $\phi$ being the azimuthal angle around the beam pipe. The pseudorapidity is defined in terms of the polar angle $\theta$ as $\eta=-\ln\tan(\theta/2)$.} measured offline less than 2.5).
The comparison between the trigger and offline tracks is done by associating the tracks within a cone of $\Delta R<0.05$\footnote{$\Delta R = \sqrt{(\Delta\eta)^2 + (\Delta\phi)^2}$, $p_T=$transverse momentum}.

\subsection{Muons}

The performance of the muon trigger selecting muon candidates with a minimum transverse momentum of \unit{10}{\giga\electronvolt} is presented.
Due to the \unit{10}{\giga\electronvolt} threshold on the muon transverse momentum, the same requirement is applied to the offline tracks.

\begin{figure}[htb]
  \centering
  \begin{subfigure}[htb]{0.47\textwidth}
      \centering
      \includegraphics[width=\textwidth]{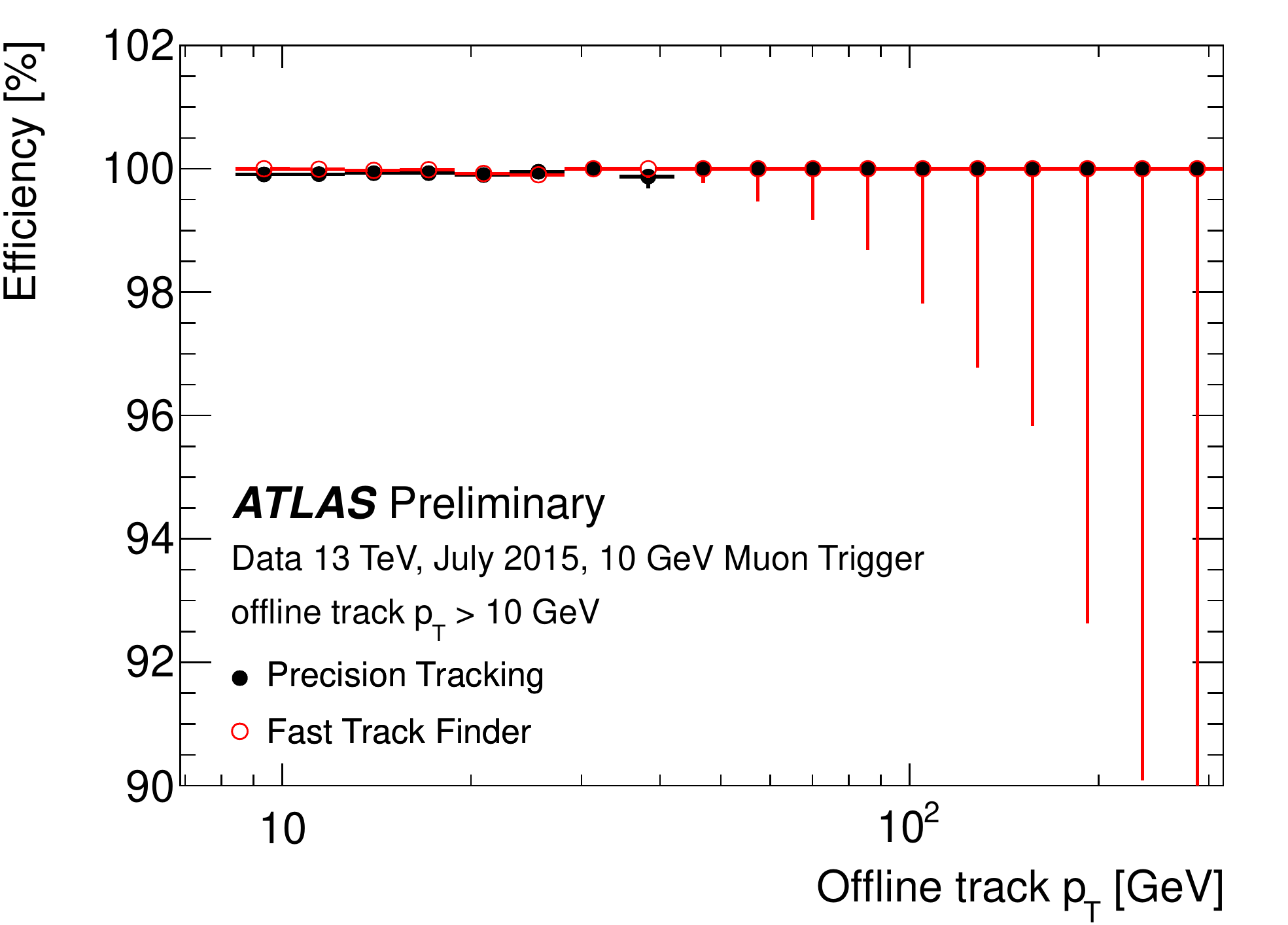}
      \caption{}
  \end{subfigure}
  ~
  \begin{subfigure}[htb]{0.47\textwidth}
      \centering
      \includegraphics[width=\textwidth]{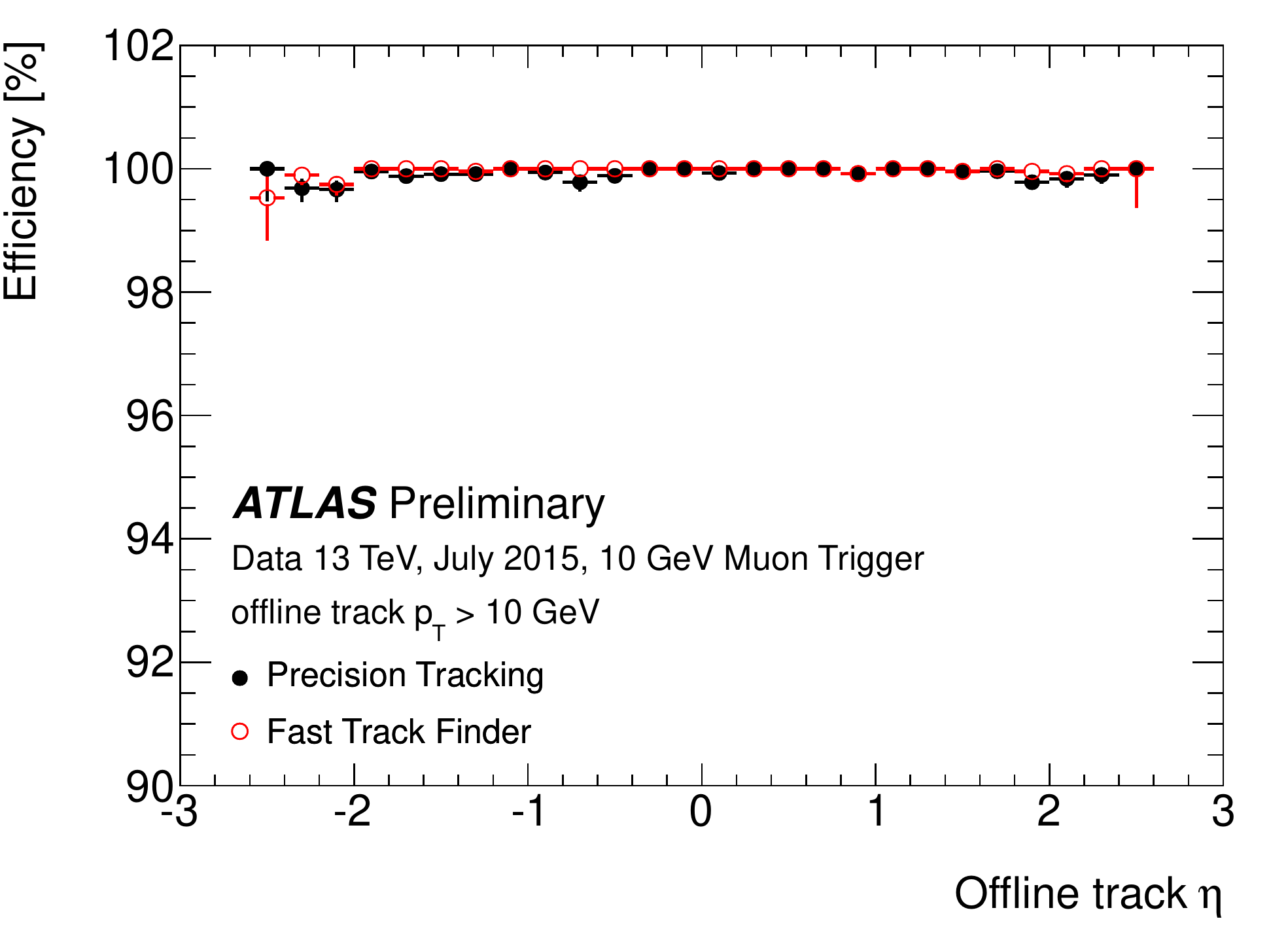}
      \caption{}
  \end{subfigure}
  \caption{The efficiency of the HLT inner detector tracking algorithms for a \unit{10}{\giga\electronvolt} muon trigger with respect to the offline track reconstruction as a function of (a) the transverse momentum and (b) the pseudorapidity. Taken from Ref.~\cite{2015TrigPerf}.}
  \label{fig:MuEff}
\end{figure}

The efficiency of the HLT inner detector tracking algorithms on these events can be seen in the plots in Fig.~\ref{fig:MuEff} as a function of the transverse momentum and the pseudorapidity respectively.
Track reconstruction efficiencies are found to be high across the full range, with no dependency on either of these track parameters.

The resolution of muons found with the HLT inner detector tracking algorithms has also been investigated both for the track pseudorapidity and the transverse impact parameter as a function of pseudorapidity.
\begin{figure}[htb]
  \centering
  \begin{subfigure}[htb]{0.47\textwidth}
      \centering
      \includegraphics[width=\textwidth]{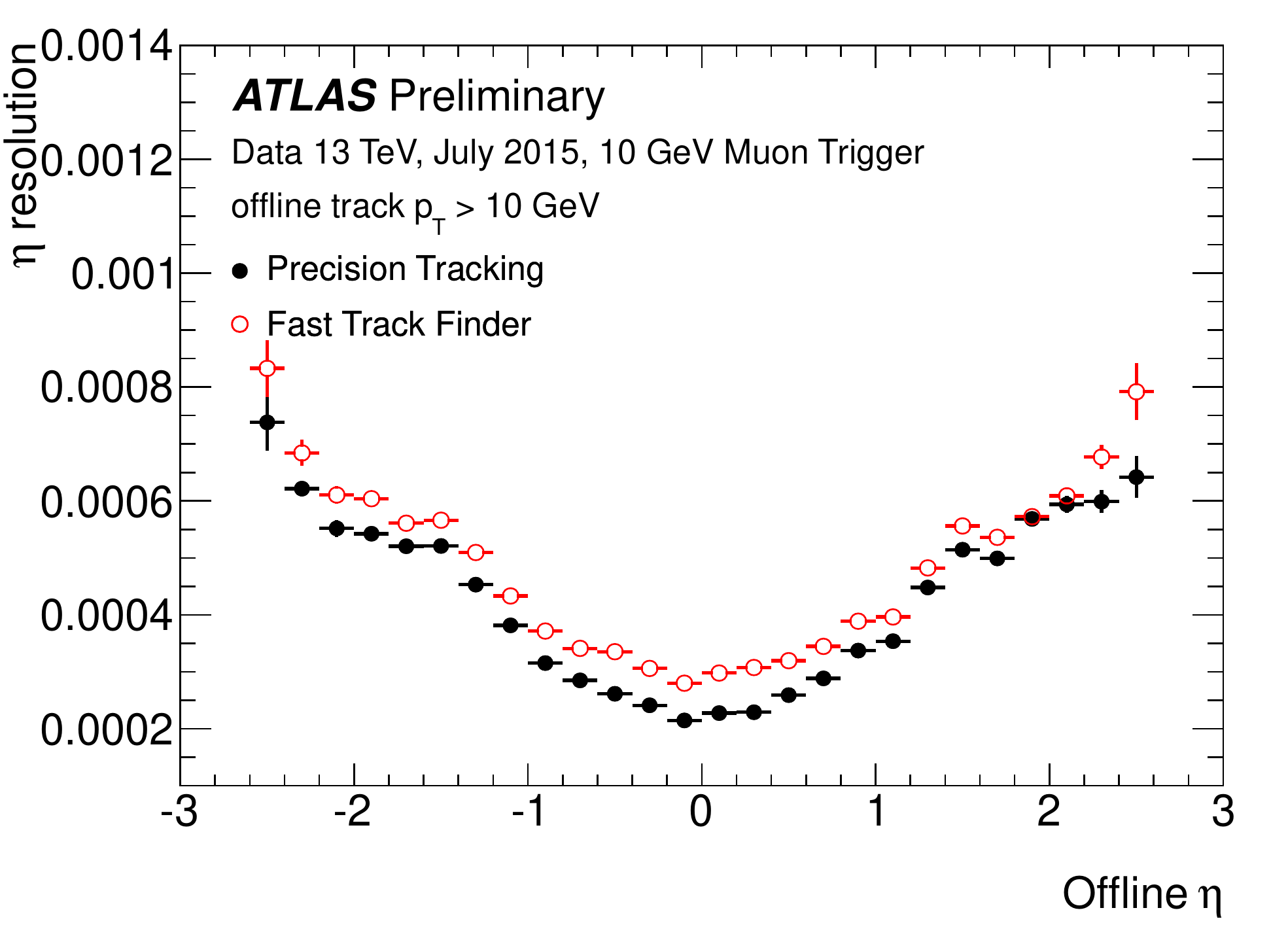}
      \caption{}
  \end{subfigure}
  ~
  \begin{subfigure}[htb]{0.47\textwidth}
      \centering
      \includegraphics[width=\textwidth]{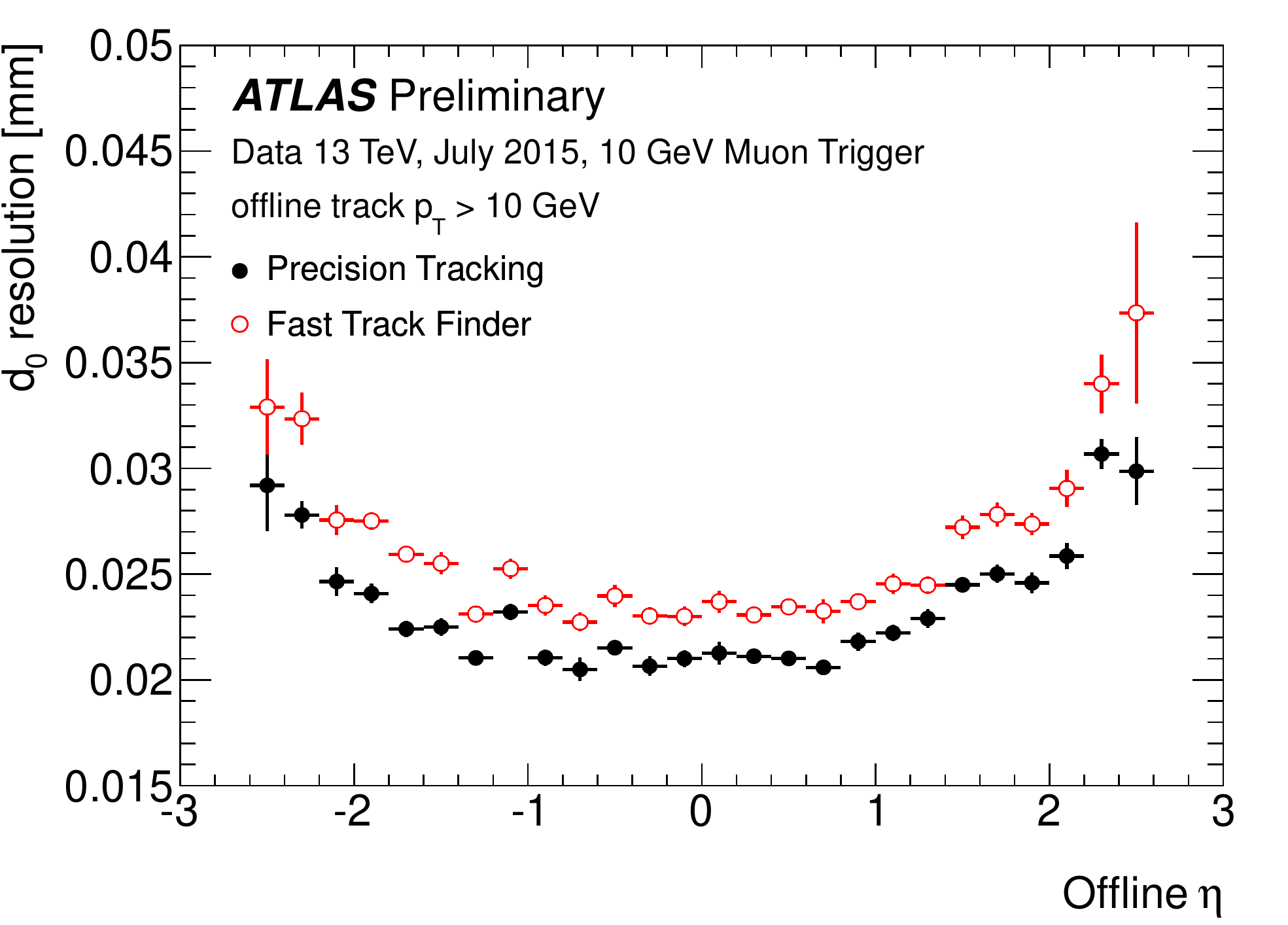}
      \caption{}
  \end{subfigure}
  \caption{The resolution of (a) the pseudorapidity and (b) the transverse impact parameter of muons found with the HLT inner detector tracking algorithms for a \unit{10}{\giga\electronvolt} muon trigger with respect to those found with the offline track reconstruction as a function of pseudorapidity. Taken from Ref.~\cite{2015TrigPerf}.}
  \label{fig:MuRes}
\end{figure}
These resolutions can be seen in Fig.~\ref{fig:MuRes}.
The resolutions are very good over the full range of pseudorapidity.
For both parameters the resolution is best at low absolute pseudorapidity, this is caused by the geometric limitations of the detector as the pseudorapidity increases.
The precision tracking stage of the trigger consistently improves the resolution over these parameters with respect to the fast track finder, as expected.

\subsection{Electrons}

The performance of the electron trigger is studied using the same data as for the muon trigger but with a different selection.
The chosen electron trigger has a threshold of \unit{24}{\giga\electronvolt} for the transverse momentum of the electron, while the offline tracks are required to have a transverse momentum of at least \unit{20}{\giga\electronvolt}.

\begin{figure}[htb]
  \centering
  \begin{subfigure}[htb]{0.47\textwidth}
      \centering
      \includegraphics[width=\textwidth]{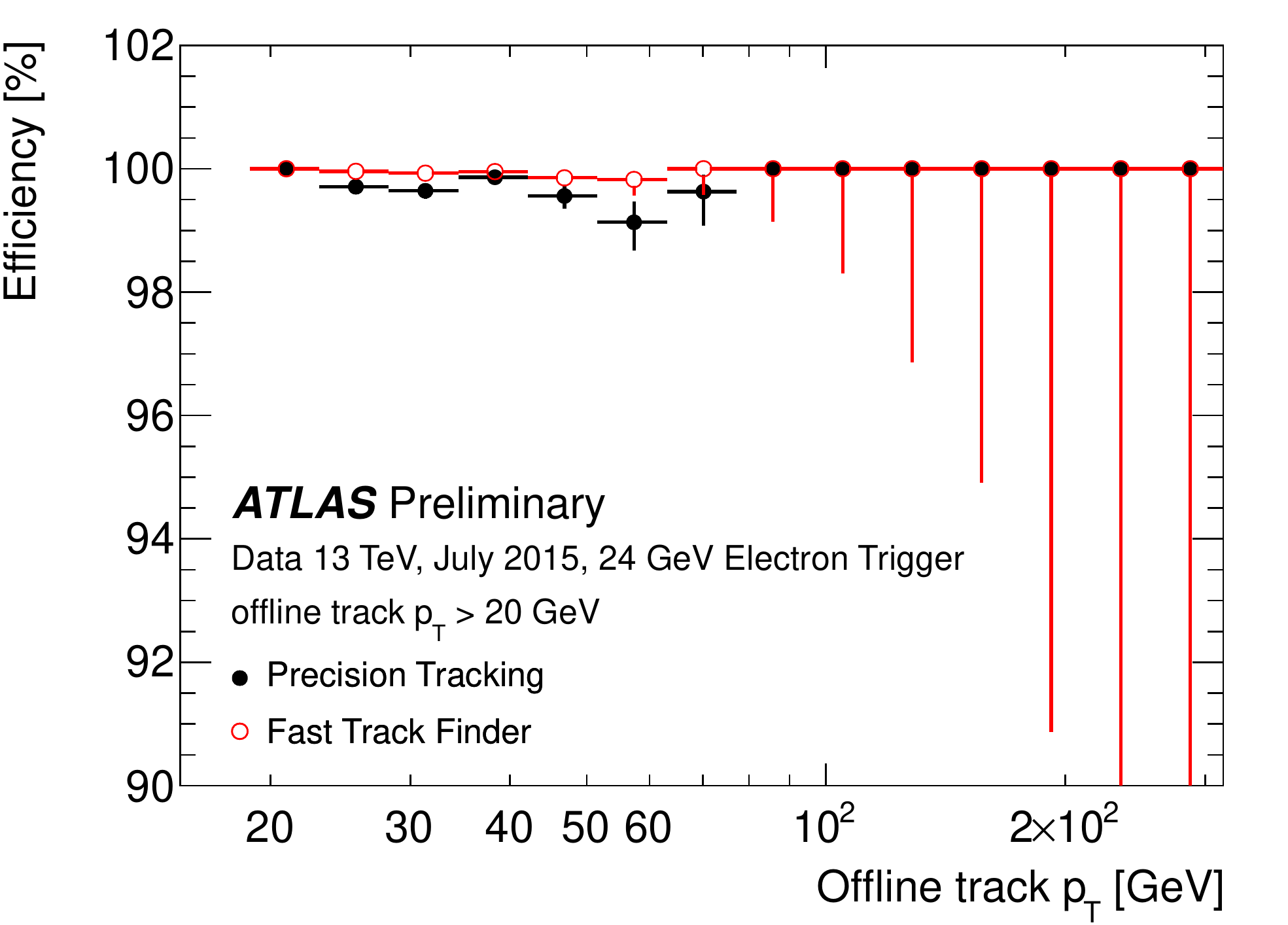}
      \caption{}
  \end{subfigure}
  ~
  \begin{subfigure}[htb]{0.47\textwidth}
      \centering
      \includegraphics[width=\textwidth]{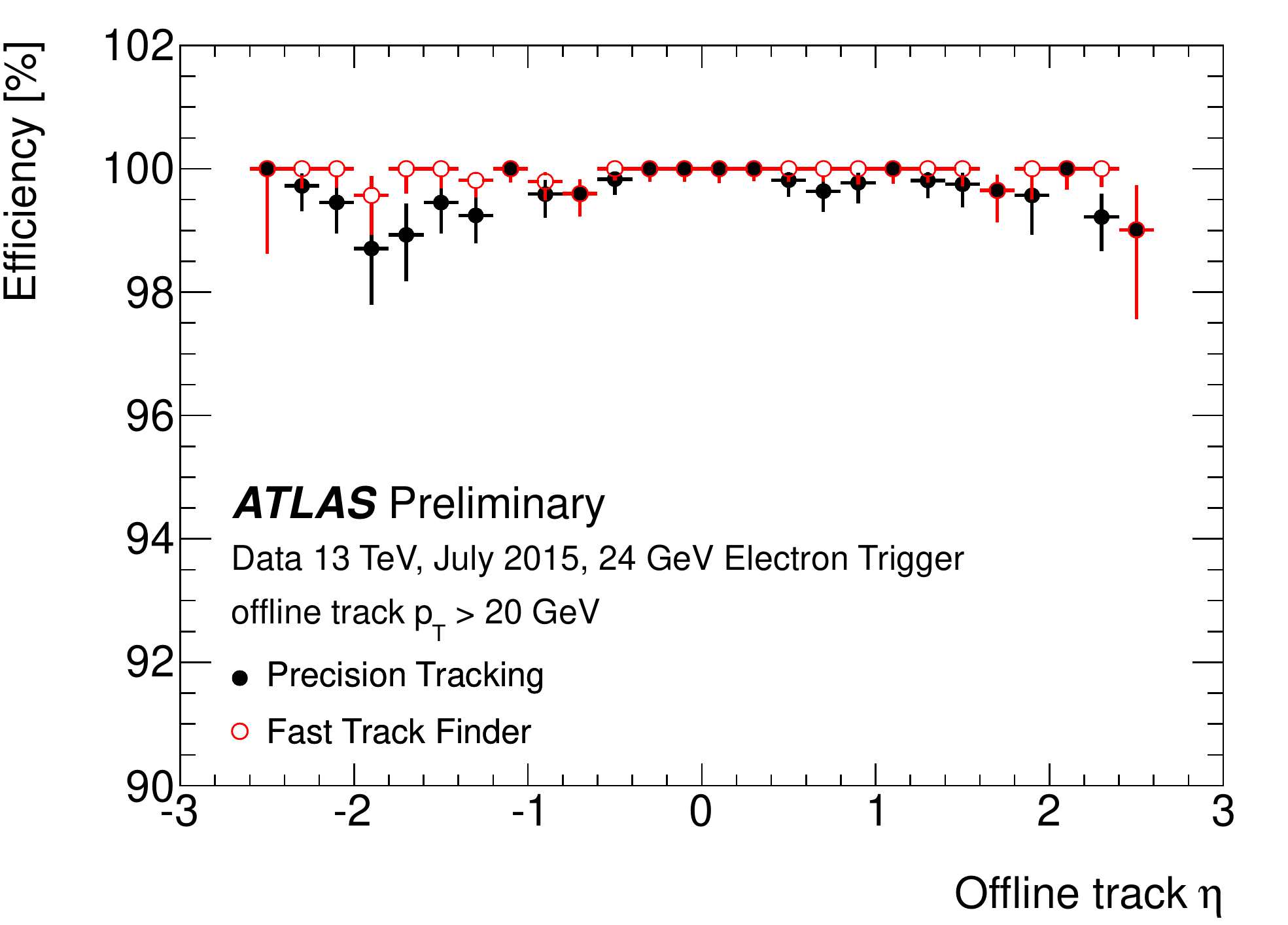}
      \caption{}
  \end{subfigure}
  \caption{The efficiency of the HLT inner detector tracking algorithms for a \unit{24}{\giga\electronvolt} electron trigger with respect to the offline track reconstruction as a function of (a) the transverse momentum and (b) the pseudorapidity. Taken from Ref.~\cite{2015TrigPerf}.}
  \label{fig:ElEff}
\end{figure}

The efficiencies for the HLT inner detector tracking algorithms run in this trigger can be seen in Fig.~\ref{fig:ElEff} for the transverse momentum and the pseudorapidity.
Triggering on electrons is expected to have slightly worse performance than muons due to electrons undergoing bremsstrahlung leading to energy losses in flight.
In offline reconstruction there are more sophisticated corrections to account for these energy losses which are too time consuming to use online.
The effect of this can be seen by the slightly lower average efficiency.
The efficiency appears flat with respect to transverse momentum.
There is a small decrease in efficiency at large $|\eta|$, but it remains above $99\%$ for the full range of tracks observed.

\section{Conclusions}

For Run 2 the data taking conditions have changed substantially leading to a much higher rate of processes considered by the trigger.
The whole trigger system has been revamped, as part of this the high level trigger has been merged into a single stage.
The inner detector trigger software has been redesigned to remove duplication of data preparation and to seed precision tracking from the initial fast tracking stage.
The changes, along with improvements in the offline algorithms, mean the time taken per event has been reduced by almost a factor of three whilst maintaining excellent efficiency.
It has been observed that there is once more excellent performance from the inner detector trigger in the first data collected in Run 2.
In the future, the algorithms will continue to be improved and tuned and the efficiency studied in greater detail.
The FTK hardware is due to be installed and commissioned in 2016 providing track seeds for the HLT, improving the speed of the inner detector trigger even further.

\FloatBarrier

\end{document}